\documentclass[twocolumn,pra,amsmath,letterpaper,floatfix,showpacs]{revtex4}
\usepackage{graphicx}
\usepackage{amsmath}
\usepackage{amssymb}
\usepackage{color}

\newcommand{\Otwo}{{O$_{2}$ }}
\newcommand{\Ntwo}{{N$_{2}$ }}

\begin{document}

\title{Sound emission from the gas of molecular superrotors}

\author{A. A. Milner, A. Korobenko, V. Milner}
\date{\today}

\affiliation{Department of  Physics \& Astronomy, The University of British Columbia, Vancouver, Canada}

\begin{abstract}
We use an optical centrifuge to deposit a controllable amount of rotational energy into dense molecular ensembles. Subsequent rotation-translation energy transfer, mediated by thermal collisions, results in the localized heating of the gas and generates strong sound wave, clearly audible to the unaided ear. For the first time, the amplitude of the sound signal is analyzed as a function of the experimentally measured rotational energy. The proportionality between the two experimental observables confirms that rotational excitation is the main source of the detected sound wave. As virtually all molecules, including the main constituents of the atmosphere, are amenable to laser spinning by the centrifuge, we anticipate this work to stimulate further development in the area of photo-acoustic control and spectroscopy.
\end{abstract}

\maketitle

Interaction of intense laser radiation with dense gases may strongly affect their optical and hydrodynamic properties. Nonlinear change of refractive index gives rise to filamentation (recent reviews of this broad topic can be found in \cite{Couairon2007,Berge2007}), whereas spatially localized heating results in the formation of acoustic waves\cite{Yu2003, Kartashov2006, Clough2010, Kiselev2011, Wahlstrand2014, Wu2014} and long-lived gas density perturbations\cite{Cheng2013, Jhajj2014, Zahedpour2014}. Photo-acoustic vibrational Raman spectroscopy has been successfully utilized for chemical detection and trace analysis\cite{Siebert1980, Melchior1998}.

In the limit of laser intensities approaching and exceeding $10^{14}$ W/cm$^{2}$, i.e. above the typical ionization threshold, local gas heating due to the re-scattering of hot electrons in laser-induced plasma filaments is well understood\cite{Filin2009}. Acoustic waves associated with this heating mechanism have been observed and studied\cite{Yu2003, Clough2010, Wu2014}. In the case of weaker non-ionizing laser pulses, rotational excitation of molecules, rather than their ionization, has been suggested as the primary mechanism of depositing energy to the molecular media\cite{Kartashov2006, Kiselev2011, Cheng2013}. Recent experiments employing time-resolved double-pulse detection schemes have provided convincing evidence of energy conversion from molecular rotation to sound\cite{Schippers2011, Zahedpour2014}.

The above mentioned studies of rotational photo-acoustics are based on the impulsive excitation of molecular rotation by a single femtosecond laser pulse or a short series of up to four pulses. These methods are limited to relatively low levels of rotational excitation with molecules gaining only a few units of angular momentum ($2\hbar$). At this limit, the amount of rotational energy deposited by the laser field is rather hard to control. As a result, the direct connection between the rotational energy and the amplitude of the sound wave has not been established beyond the perturbative limit. The latter is characterized by a weak nonlinear absorption through a single two-photon Raman transition, for which the expected quadratic scaling of the acoustic amplitude with the pulse intensity has been reported \cite{Kartashov2006, Kiselev2011}.

In this work, we employ the technique of an optical centrifuge\cite{Karczmarek1999, Villeneuve00}, recently developed in our group for applications in dense gas media\cite{Korobenko14a}, as a powerful tool for controlled rotational excitation of molecular ``superrotors'' - molecules carrying an extremely large amount of angular momentum, e.g. $>70 \hbar$ for nitrogen and $>120 \hbar$ for oxygen. Utilizing the tunability of the centrifuge\cite{Milner15a}, we vary the amount of rotational energy deposited  into the gas of \Ntwo or \Otwo molecules while recording the acoustic signal produced by the superrotors. The latter is clearly audible to the unaided ear due to the ultra-high rotational energies exceeding 2.5 eV per oxygen superotor. Our state-resolved detection method, based on coherent Raman scattering, enables us to analyze, for the first time, the dependence of the recorded sound wave amplitude on the degree of rotational excitation, and confirm the theoretically predicted scaling.
\begin{figure}[b]
\centering\includegraphics[width=1\columnwidth]{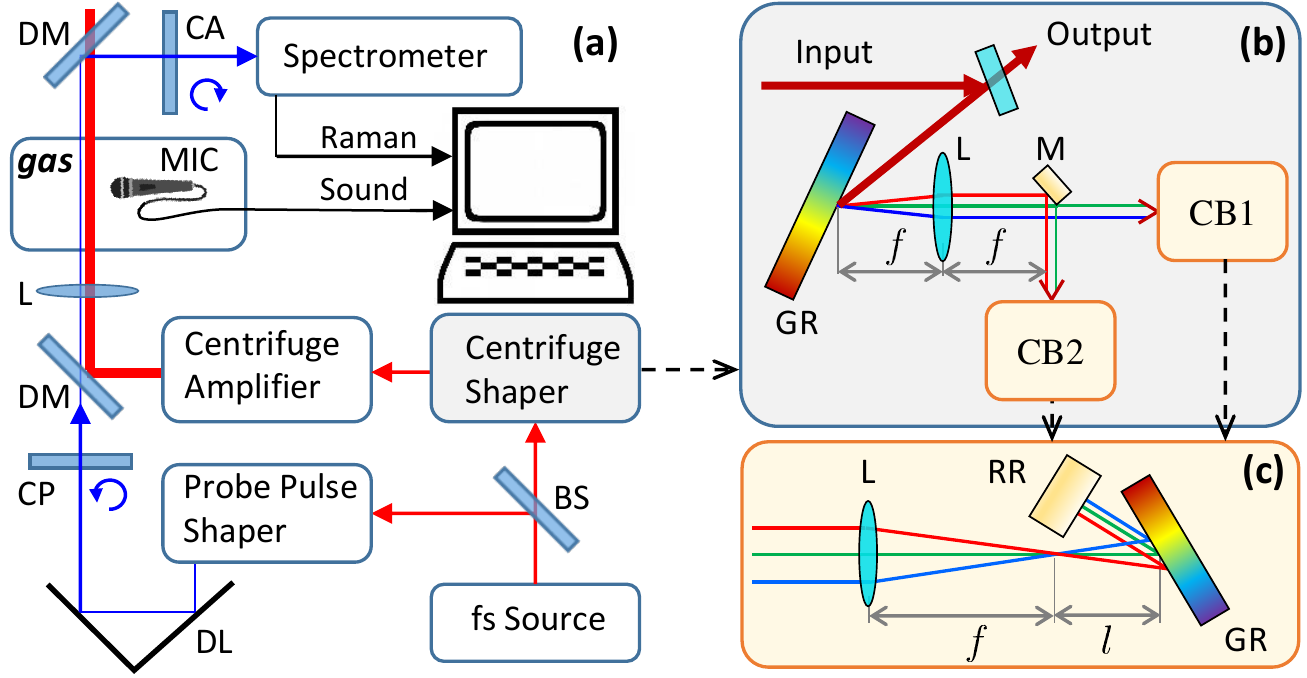}
\caption{(\textbf{a}): Experimental set up. BS: beam splitter, DM: dichroic mirror, CP/CA: circular polarizer/analyzer, DL: delay line, L: lens, MIC: Microphone. (\textbf{b}): Centrifuge shaper. GR: grating, M: pick-off mirror in the Fourier plane of lens L of focal length $f$, CB1 and CB2: two ``chirp boxes'', schematically shown in panel (\textbf{c}), where RR denotes a retro-reflector and length $l$ controls the applied frequency chirp.}
\label{Fig-Setup}
\end{figure}

The description of our experimental setup, schematically shown in Fig.\ref{Fig-Setup}(\textbf{a}), can be found in Ref.\cite{Milner14b}. Briefly, a beam of femtosecond pulses from an ultrafast laser source (spectral full width at half maximum (FWHM) of 30 nm) is split in two parts. One part is sent to the ``centrifuge shaper'' (Fig.\ref{Fig-Setup}(\textbf{b,c})) which converts the input laser field into the field of an optical centrifuge according to the original recipe of Karczmarek \textit{et al.} \cite{Karczmarek1999}. The centrifuge shaper is followed by a home built Ti:Sapphire multi-pass amplifier boosting the pulse energy to 30 mJ. FWHM diameter of the centrifuge beam in the focal plane is 90 $\mu $m, which results in the maximum peak intensity of $2.2\times10^{12}$ W/cm$^{2}$. Importantly, this is about two orders of magnitude below the ionization threshold of nitrogen and oxygen\cite{Kartashov2006, Schippers2011}, which ensures the non-ionizing mechanism of the laser-induced gas heating studied in this work.

\begin{figure}[b]
\centering\includegraphics[width=1\columnwidth]{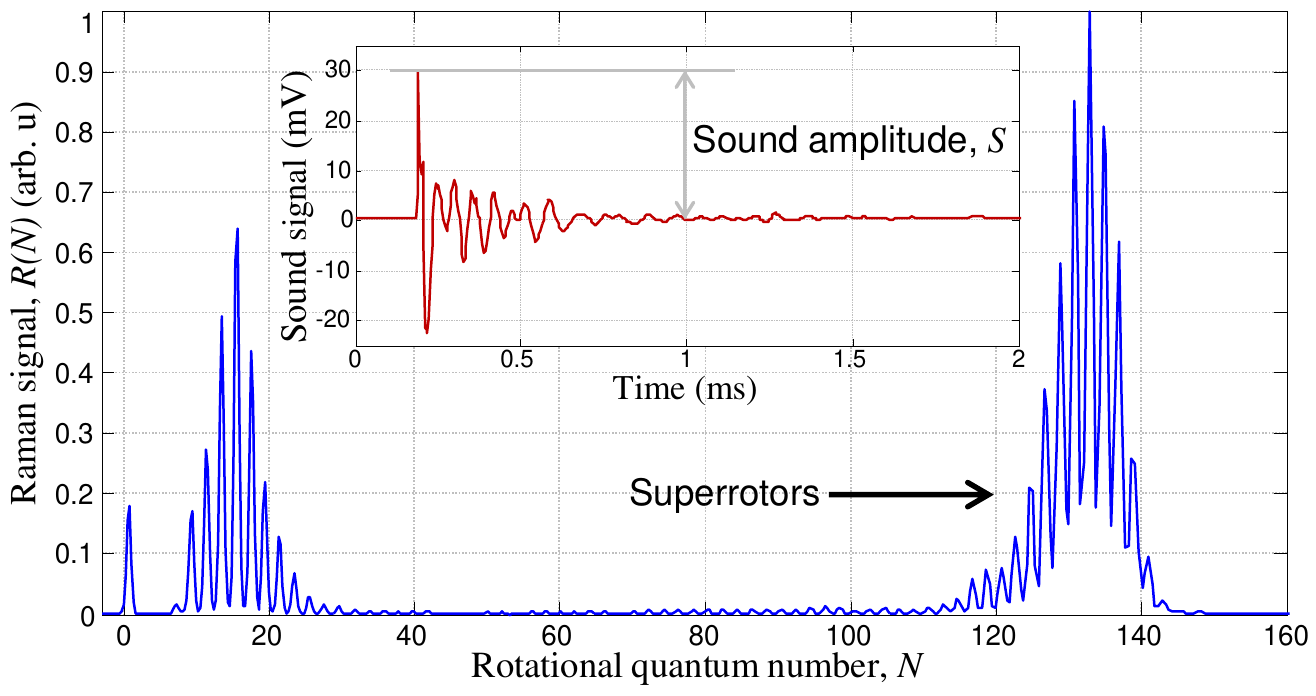}
\caption{Typical rotational Raman spectrum of oxygen superrotors measured with a probe pulse delayed by 200 ps with respect to the centrifuge pulse (see text for the description of various components). Inset: an example of the acoustic signal recorded in centrifuged ambient air.}
\label{Fig-Signals}
\end{figure}

The second beam (probe, central wavelength of 398.2 nm) passes through the standard $4f$ Fourier pulse shaper employed for narrowing the spectral width of probe pulses down to 3.75 cm$^{-1}$ (FWHM). The centrifuge-induced rotational coherence results in the Raman frequency shift of the probe field, discussed in detail in our previous work\cite{Korobenko14a}. An example of the experimentally detected probe spectrum $R(N)$, recorded with probe pulses delayed by 200 ps with respect to the centrifuge propagating in oxygen, is shown in Fig.\ref{Fig-Signals}. Horizontal scale has been converted from the measured energy shift $\Delta E$ to the rotational quantum number $N$ according to $\Delta E= E_{N+2} - E_{N}$, where $E_{N}=B N (N+1) - D N^2 (N+1)^2$ is the rotational energy, with $B$ and $D$ being the rotational and centrifugal constants of O$_{2}$, respectively. The state-resolved spectrum consists of individual rotational peaks which are grouped into the following three components. The $N=0$ line corresponds to the frequency-unshifted probe light. The wave packet centered around $N=15$ corresponds to the molecules which were too hot to follow the centrifuge and spilled out of it at the beginning of the spinning process. A set of farther lines around $N=133$ represents the signal from the centrifuged molecules. By shaping the spectrum of the centrifuge pulse, the latter wave packet can be centered anywhere between $N=15$ and 133.

The acoustic signal is detected by a microphone positioned perpendicularly to the centrifuge beam 6 cm away from its focal spot. The sensitivity of the microphone is $-31$ dBV ($\approx 28$ mV/Pa) at 1 KHz under ambient conditions, and its output is connected directly to the computer's analog-to-digital converter with no further amplification. An example of the recorded acoustic signal, averaged over 100 laser pulses, is shown in the inset to Fig.\ref{Fig-Signals}. Frequency bandwidth of the microphone is not sufficiently broad for resolving the effectively instantaneous pressure burst, which results in the decaying oscillatory waveform. Hereafter, we refer to the maximum value of this waveform as the amplitude of the acoustic signal $S$.

\begin{figure*}[t]
\centering\includegraphics[width=2\columnwidth]{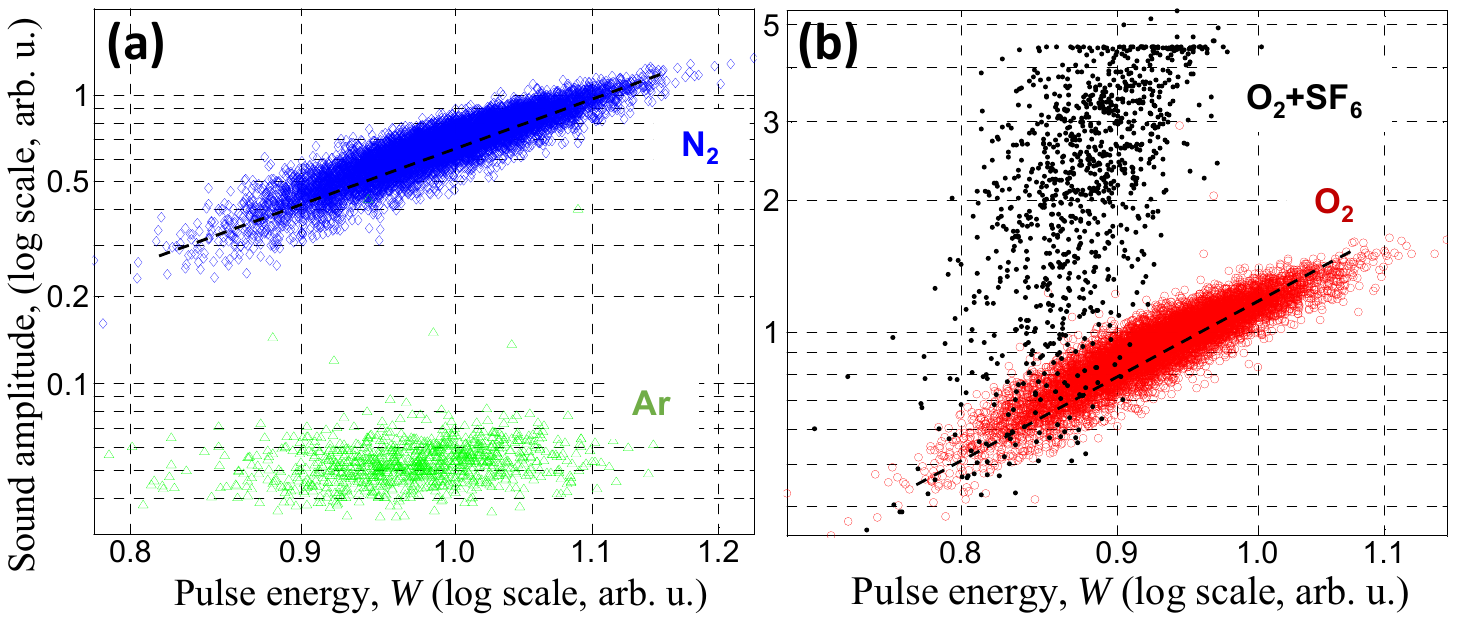}
\caption{Amplitude of the recorded sound as a function of the energy of centrifuge pulses, plotted on a \textit{log-log} scale. Each data set consists of 10,000 points. (\textbf{a}) Typical acoustic signal from the centrifuged gas of nitrogen molecules (blue diamonds) is compared to the sound generated by the centrifuge in pure argon at the same pressure of 95 kPa (green triangles). (\textbf{b}) Acoustic response of the centrifuged oxygen with and without a small admixture of SF$_{6}$ molecules (black dots and red circles, respectively). Black dashed lines in both panels show fits to power-law scaling.}
\label{Fig-EnergyScaling}
\end{figure*}

The acoustic response of various gases to the field of an optical centrifuge is shown in Fig.\ref{Fig-EnergyScaling} as a function of the laser pulse energy. We first notice the difference between the sound amplitude detected in the ensemble of nitrogen superrotors and in pure argon gas under the same pressure and temperature, clear from Fig.\ref{Fig-EnergyScaling}(\textbf{a}). Even though the ionization energies of the two molecules are almost the same, 15.58 eV for \Ntwo and 15.76 eV for Ar\cite{NIST}, the acoustic signal in nitrogen is an order of magnitude stronger. This confirms our earlier intensity-based argument about the non-ionizing nature of the sound waves recorded in the gas of superrotors.

To examine this hypothesis further, we intentionally introduce ionization by adding a small amount of SF$_{6}$ molecules into the gas of \Otwo. Being a spherical top, SF$_{6}$ is immune to centrifuge spinning but causes the formation of a clearly visible plasma channel. As can be seen in Fig.\ref{Fig-EnergyScaling}(\textbf{b}), when ionization is the dominant reason for the gas heating, an amplitude of the generated sound wave shows very different energy dependence than that observed in pure oxygen. Note that the stronger sound signal in plasma can be attributed to the higher number of ionized molecules in comparison to molecular superrotors. The non-ionizing mechanism of the recorded sound is also evident from the following test. We modify the laser field in such a way as to keep the direction of its vector of linear polarization constant, in contrast to the spinning polarization of the centrifuge pulse, while retaining the pulse duration and peak intensity. We observe that together with the forced accelerated molecular rotation, the acoustic signal disappears beyond our sensitivity level.

Shown by black dashed lines in Fig.\ref{Fig-EnergyScaling}, an average sound amplitude from both nitrogen and oxygen superrotors exhibits power-law scaling with pulse energy. The exact power is not universal and typically varied between 4 and 5 in our experiments with nitrogen, and between 3 and 4 with oxygen. According to the theory of laser-induced pressure waves in dense media\cite{Heritier1983}, as well as numerical simulations of hydrodynamic expansion\cite{Wahlstrand2014}, the sound wave amplitude is expected to scale linearly with the amount of laser energy deposited in the sample. In the case of an impulsive rotational excitation, most of the energy is transferred via a single two-photon Raman transition, suggesting a quadratic dependence on the pulse energy\cite{Kartashov2006}, which has been recently verified experimentally\cite{Kiselev2011, Zahedpour2014}. In contrast, an adiabatic excitation by the centrifuge involves multiple transitions up the rotational ladder, which explains our finding of much steeper scaling of the acoustic signal with laser power.

Because the efficiency of the centrifuge spinning depends on the molecular parameters, such as the moment of inertia and the polarizability anisotropy, this scaling is different for different molecules (compare \Ntwo and \Otwo in panels (\textbf{a}) and (\textbf{b}) of Fig.\ref{Fig-EnergyScaling}, respectively). The scatter of data points within each set is due to the fluctuations of the spatial and spectral shape of centrifuge pulses, which cause shot-to-shot variations in the centrifuge efficiency.

\begin{figure*}[t]
\centering\includegraphics[width=2\columnwidth]{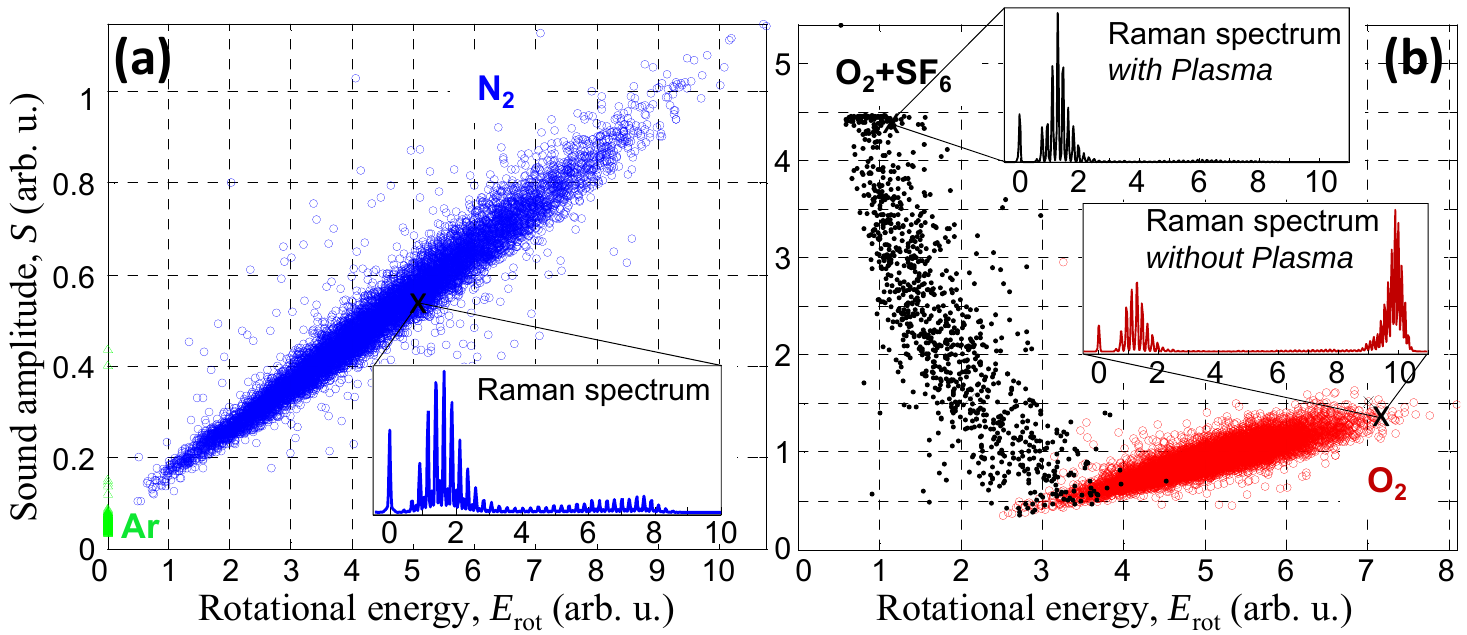}
\caption{Amplitude of the recorded sound as a function of the rotational energy deposited in the gas sample. (\textbf{a}) Acoustic response from the centrifuged gas of nitrogen molecules (blue circles) is compared to the sound generated by the centrifuge in pure argon at the same pressure of 95 kPa (green triangles). (\textbf{b}) Acoustic response from the centrifuged oxygen with and without a small admixture of SF$_{6}$ molecules (black dots and red circles, respectively). All insets show Raman spectra corresponding to the data points marked with black crosses and plotted as a function of rotational frequency in THz.}
\label{Fig-RamanScaling}
\end{figure*}

We eliminate the uncertainty in the degree of rotational excitation by sending a weak probe pulse behind each centrifuge pulse and recording the rotational Raman spectrum simultaneously with the corresponding photo-acoustic signal. This enables us to evaluate the amount of total rotational energy deposited in the molecular sample as a sum of rotational populations $P_{N}$ weighted with the corresponding rotational energy $E_{N}$ and taken over the rotational states $|N\rangle$ populated by the centrifuge:
\begin{equation}\label{Eq-Energy}
E_\text{rot} = \sum_{N} P_{N} E_{N}.
\end{equation}
A crude estimate of populations $P_{N}$ can be obtained from the measured Raman spectrum. Indeed, the intensity of each Raman peak is proportional to the square of the centrifuge-induced coherence $\rho _{N,N+2}$ between the two resonant states $|N\rangle$ and $|N+2\rangle$. For $N$'s extending beyond the thermally populated states $N_\textrm{th}$ (i.e. $\gtrsim20$ for \Ntwo and $\gtrsim30$ for \Otwo), an optical centrifuge excites pure quantum states of maximum coherence, $|\rho _{N,N+2}|^{2} = P_{N} P_{N+2}$. If the created rotational wave packet is rather broad with a smoothly changing envelope, which is the case here, one may approximate $P_{N} \approx P_{N+2}$ and correspondingly $|\rho _{N,N+2}|^{2} \approx P_{N}^{2}$. Hence, the deposited rotational energy can be estimated from the measured Raman spectrum as
\begin{equation}\label{Eq-ApproxEnergy}
E_{\text{rot}} \approx \sum_{N>N_\text{th}} \sqrt {R_{N}} E_{N}.
\end{equation}

In Figure \ref{Fig-RamanScaling}, the dependence of the sound amplitude on the deposited rotational energy, extracted from the observed Raman spectra and calculated with Eq.\ref{Eq-ApproxEnergy}, is shown for the same experimental runs as in Fig.\ref{Fig-EnergyScaling}. For nitrogen and oxygen superrotors (left and right panels, respectively), the scaling of the average photo-acoustic signal with $E_\text{rot}$ appears to be linear, in agreement with the hydrodynamic theory of heat transfer in dense media. The smooth envelope shape of the rotational wave packets in nitrogen, shown in the inset to Fig.\ref{Fig-RamanScaling}(\textbf{a}), together with the anticipated linear scaling of $S(E_\text{rot})$, confirms the validity of the approximation we used to arrive at Eq.\ref{Eq-ApproxEnergy}. In oxygen, the created wave packets are typically not as broad and smooth as in nitrogen (lower inset in Fig.\ref{Fig-RamanScaling}(\textbf{b}), but the observed scaling is still linear with $E_\text{rot}$.

To emphasize the difference in the acoustic response in the presence of plasma, the same plot shows the experimentally detected distribution $S(E_\text{rot})$ in the mixture of \Otwo with SF$_{6}$. Whenever the plasma channel is formed, the wavefront of the centrifuge beam is perturbed, which leads to the loss of coherent rotational excitation and the disappearance of superrotors, as demonstrated by the Raman spectrum in the upper inset to Fig.\ref{Fig-RamanScaling}(\textbf{b}). An increase in the sound loudness is reflected by the observed anti-correlation between the two quantities. We also verified that including the low-frequency part of the Raman spectra ($N<N_\text{th}$), corresponding to hot molecules lost from the centrifuge, did not affect the observed linear scaling, most probably due to the small amount of rotational energy carried by those low-$N$ rotors.

To summarize, we used an optical centrifuge to deposit a controllable high amount of rotational energy in a dense molecular gas, and studied its acoustic response to such rotational excitation. The predicted linear scaling of the recorded sound amplitude with the amount of deposited rotational energy has been observed in the ensemble of molecular superrotors, well below the ionization threshold, i.e. when the acoustic effects related to plasma formation are negligible. High sound levels, induced by the centrifuge pulses and audible to the unaided ear, suggest that the centrifuge technique may be useful in the broad field of photo-acoustics.

This work has been supported by the grants from CFI, BCKDF and NSERC. We thank Ilya Sh. Averbukh for useful discussions.


\end{document}